\documentclass[aps,pra,epsfigure,twocolumn,showpacs]{revtex4}
\usepackage{dcolumn}    
\usepackage{bm} 
\usepackage{graphicx}
\usepackage{amsmath}    
\usepackage{latexsym}
\usepackage{amsfonts}   
\usepackage{amssymb}
\usepackage{array}      
\usepackage{epsfig}
\usepackage{epstopdf}

\newcommand{\ket}[1]{\left\vert#1\right\rangle}

\newcommand{\bra}[1]{\left\langle#1\right\vert}

\newcommand{\gs}{\text{gs}}

\begin{document}
\title{Multipartite non-locality in a thermalized Ising spin-chain}
\author{Steve Campbell and Mauro Paternostro}
\affiliation{School of Mathematics and Physics, Queen's University, Belfast BT7 1NN, United Kingdom}

\begin{abstract}
We study multipartite correlations and non-locality in an isotropic Ising ring under transverse magnetic field at both zero and finite temperature. We highlight parity-induced differences between the multipartite Bell-like functions used in order to quantify the degree of non-locality within a ring state and reveal a mechanism for the passive protection of multipartite quantum correlations against thermal spoiling effects that is clearly related to the macroscopic properties of the ring model. 
\end{abstract}
\date{\today} 
\pacs{03.67.Bg,03.65.Ud}
\maketitle

\section{Introduction}
\label{Intro}
Correlated systems are fundamental in developing and exploiting the possibilities arising from quantum mechanics. A premier role in such an endeavor is played by chains of interacting quantum spins~\cite{sugato}, in light  of their wide applicability to many physical settings and their ability to simulate the behavior of strongly correlated systems in condensed matter~\cite{heisen,abel}. They are a key ingredient in many distributed quantum computation and communication protocols~\cite{bose2}. If we are truly to benefit from the potential of such resources, a thorough understanding of the nature of such correlations is essential. In this respect, it is particularly important to gather information on the distribution and amount of genuine multipartite quantum correlations in one of such systems so as to be able to tailor specific protocols to the available structure of entanglement among the parties of a chain or lattice. 

Entanglement and non-locality are the key figures of merit for non-classical correlations. Already a good understanding exists for simple bipartite entangled systems and there is much research in the direction of multipartite entangled states~\cite{HHHH}. Equally, non-locality is truly at the forefront of current research as the premier manifestation of true deviation  of the behavior of a system from classicality~\cite{popescu, exp, brunner}. Non-locality embodies a stronger type of correlation than entanglement and the development of tools to detect, quantify and characterize it is moving swiftly in the field. Of course one should be mindful of how any such system can be experimentally generated and utilized and of the fact that unavoidable interactions with a surrounding environment result in the loss of such strong quantum correlations. Methods to control the affect of these unwanted interactions (resulting in either losses or pure decoherence) have been proposed, ranging from bang-bang techniques~\cite{bang} to the active use of the quantum Zeno effect~\cite{zeno}. All such methods, to the best of our knowledge, require either heavy influences on the system at hand or expensive additional resources for entanglement protection. The generation of non-classical multipartite correlated states is already a formidable task: if loss/decoherence-tolerant methods for their production have to be in order, this would necessarily imply a frequently prohibitively escalating experimental cost. It should thus be evident that a scenario where highly quantum correlated states are produced as the ground state of a given many-body interacting model and {\it naturally} maintained while in the proximity of environmental effects, is endowed with extreme physical appeal.

In this paper we present a study of the Ising model of spin-1/2 particles (qubits) in a transverse magnetic field. This model has attracted wide attention due to its exactly solvable nature and the attractive natural properties of its ground states~\cite{buzek, kendon,iranians}. We go significantly further than the available literature on this model so far by addressing the case of genuine multipartite quantum correlations and their nature. We examine the zero temperature dynamics of entanglement and non-locality highlighting by means of an efficient hybrid approach for the determination of multi-site correlation functions. The finite temperature regime is then addressed and we show the ability to control and preserve the non-locality present in the system with minimal additional effort. We show the key ingredient in the model is the presence and manipulation of the magnetic field.

The remainder of the paper is organized as follows. In Sec.~\ref{Secmodel} we introduce the physical model for spin-spin coupling considered in this work and sketch the technique required for its full diagonalization. Sec.~\ref{Sectools} discusses the technical tools used in our investigation for the quantification of multipartite entanglement and quantum non-locality. Sec.~\ref{nonclassical} presents the dynamics of multipartite correlations at zero temperature while Sec.~\ref{TnonZero} we consider how entanglement and non-locality are affected in the finite temperature case. Finally Sec.~\ref{conclus} summarizes our findings.

\section{The Model}
\label{Secmodel}

We begin our study by discussing features and properties of an isotropic Ising model in a transverse magnetic field. For $N$ coupled spins the Hamiltonian model reads
\begin{equation}
\label{model0}
\hat{\cal H}=-{\cal J}\sum^{N}_{n=1} \hat\sigma^{x}_{n} \otimes \hat\sigma^{x}_{n+1} + {\cal B} \sum^{N}_{n=1} \hat\sigma^{z}_{n}
\end{equation}
with ${\cal J}$ the inter-spin coupling strength and ${\cal B}$ the global magnetic field for each qubit and $\sigma^i_n$ ($i=x,y,z$) the $i$ Pauli spin operators for spin $n=1,..,N$. In order to strip our analysis from unnecessary complications, we work with a dimensionless version of the Hamiltonian defined by assuming ${\cal J}=J{\cal E}$ and ${\cal B}=B{\cal E}$ with ${\cal E}$ a common order-of-magnitude factor and $J,B$ are dimensionless parameters. We thus consider the dimensionless Hamiltonian 
\begin{equation}
\label{model}
\hat{H}\equiv\frac{\hat{\cal H}}{\cal E}=-J\sum^{N}_{n=1} \hat\sigma^{x}_{n} \otimes \hat\sigma^{x}_{n+1} + B \sum^{N}_{n=1} \hat\sigma^{z}_{n}
\end{equation}
We can readily determine the thermal state associated with this Hamiltonian using
\begin{equation}
\label{thermalstate}
\varrho(T)= e^{-\frac{\beta}{T}\hat{H}}/{{\cal Z}}
\end{equation}
where ${\cal Z}$ is the partition function and $\beta=\hbar/k_B$. Clearly each spin only interacts with its nearest neighbors and we shall assume a cyclic boundary condition such that the last spin interacts with the first, thus forming a spin-ring. It is also worth noting that we have taken the magnetic field to act orthogonal to the interaction. This particular model is very attractive due to its exactly solvable nature. The solution is presented in~\cite{buzek,mattis,osborne}, and so for completeness we simply comment on the methods employed. In order to diagonalize the Hamiltonian we use the Jordan-Wigner transformation~\cite{JW}. Firstly, we transform the Hamiltonian using a new variable $\sigma^{\pm}=1/2(\sigma^x_n \pm i\sigma^y_n)$. We can then easily move into the fermionic picture by introducing a further transformation
\begin{equation}
\hat c^\dag_n=(\hat c_n)^\dag=\otimes^{n-1}_{j=1}(-\hat{\sigma}^z_j)\hat{\sigma}^+_n~~~~\{c_n,c^\dag_m\}=\delta_{nm}
\end{equation}
with $\delta_{nm}$ the Kronecker delta, which is $1$ for $n=m$ and $0$ otherwise.
These fermionic variables then allow the Hamiltonian to be written in the form
\begin{equation}
\hat H{=}-J\sum^{N}_{n=1}(\hat c^{\dagger}_n-\hat c_n)(\hat c^{\dagger}_{n+1}+\hat c_{n+1})+2B \sum^{N}_{n=1}(\hat c_n^\dagger \hat c_n -\frac{1}{2}).
\end{equation}
Using a Fourier transform we pass to the momentum representation. The resulting Hamiltonian can then be diagonalized by means of a Bogoliubov transformation introducing new fermionic operators $\{\hat b_k,\hat b^\dag_k\}$. The latter are related to the Fourier-transformed fermionic operators $\hat{c}_{\pm k}$ through $\hat{b}_{k}=\cos(\vartheta_k/2)\hat{c}_k-i\sin(\vartheta_k/2)\hat{c}^\dag_{-k}$ and $\hat{b}^\dag_{k}=\cos(\vartheta_k/2)\hat{c}^\dag_k+i\sin(\vartheta_k/2)\hat{c}_{-k}$. This brings it into the free-fermion form
\begin{equation}
\label{free}
\hat{H}_{\text{ff}}=\sum_{k}\epsilon_k\hat{b}^\dag_k\hat{b}_k- \sum_k\epsilon_k,
\end{equation}
where ${\epsilon_k=\sqrt{J^2+B^2-2JB\cos \phi_k}}$, $\tan\vartheta_k=(-B+J\cos\phi_k)/(J\sin\phi_k)$ and ${\phi_k=\pi(2 k+1)/N}$ with $k=-N/2,..,N/2-1$ in the subsector with an even number of fermions and a slightly different definition for the odd-number case. However, in what follows we will be mostly concerned with the ground state of the Hamiltonian, which lives in the even-number sector (the appropriate expression will be used when leaving the ground state).  The constant term in Eq.~(\ref{free}) is related to $\sum_k\epsilon_k$. The ground state of the system is then the state with no population in any of the decoupled $\hat{b}_{k}$ modes, {\it i.e.} the state solves the set of conditions $\hat b_k\ket{\gs_N}=0$ simultaneously, for any allowed value of ${k}$. A standard calculation (see for details Ref.~\cite{rico}) then leads to the ground state
\begin{equation}
\ket{\gs_N}=\bigotimes_{k}[\cos\frac{\vartheta_k}{2}\ket{00}_{k,-k}+i\sin\frac{\vartheta_k}{2}\ket{11}_{k,-k}]
\end{equation}
with $\ket{0}_{\phi_k}$ ($\ket{1}_{\phi_k}$) the state with 0 (1) fermions with momentum $\phi_k$. This  approach is particularly useful when a ring with a large number of particles is considered. In this case, the direct diagonalization of Eq.~(\ref{model}), in fact, becomes computationally demanding. 
Regardless of the dimension of the ring at hand, the magnetic field proves to be key to the behavior of entanglement and non-locality within our system. In fact, when no magnetic field is present, from Eq.~(\ref{free}) we see that $\epsilon_k=J~\forall{k}$, so that the ground state has energy $-NJ$, which is the same for $\otimes^N_{n=1}\ket{+}_n$ and $\otimes^N_{n=1}\ket{-}_n$ with $\ket{\pm}_n=(\ket{0}_n\pm\ket{1}_n)/\sqrt 2$ being the eigenstates of $\hat\sigma^x_n$. In these expressions, $\{\ket{0}_n,\ket{1}_n\}$ is the computational logic basis of spin $n=1,..,N$. As Stelmachovic and Buzek proved in Ref.~\cite{buzek}, the proper ground state of the ring in these conditions can be taken as a Greenberger-Horne-Zeilinger state~\cite{GHZ} in the basis built from the tensor product of eigenstates of $\hat\sigma^x_n$. That is
\begin{equation}
\ket{\text{GHZ}^x_N}=(\ket{++..+}+\ket{--..-})/\sqrt 2,
\end{equation}
which, when written in the $\otimes^N_{n=1}\hat\sigma^z_n$ eigenbasis, always has an even number of spins in their $\ket{1}$ state. In what follows, the energy of the ground state on a system of $N$ spins is indicated as 
\begin{equation}
\Lambda_N=-\sum_k\epsilon_k.
\end{equation}

\section{Tools for studying multipartite entanglement and non-locality}
\label{Sectools}

In order to examine the dynamics of non-locality and entanglement, we need appropriate instruments to quantify them. Bipartite entanglement can be quantified by a number of well-defined measures. However for mixed, multipartite states no such measure exists~\cite{HHHH}. Here we we shall employ a suitable generalization of negativity, an entanglement measure based on the well-known Peres-Horodecki criterion~\cite{pereshorodecki}, which is necessary and sufficient for $2 \times 2$, $2 \times 3$ and $\infty \times \infty$ systems., together with an extension of Wootters concurrence~\cite{Wconc} to the case of a multipartite register of an even number of elements. 

For a two-qubit system, negativity is defined as
\begin{equation}
\label{bipartneg}
{\cal N}_2=-2\text{max} [0,\lambda_{\text{neg}}],
\end{equation}
where $\lambda_{\text{neg}}$ is the single negative eigenvalue of the partially transposed version of the density matrix describing the state of the qubits. This measure can be extended to tripartite systems by simply considering the partial transposition of the density matrices associated with appropriate bipartitions of a given system~\cite{sabin}. Let us consider three spins, $A, B$ and $C$. The entanglement of spin $A$ with spins $B$ {\it and} $C$ is
\begin{equation}
{\cal N}_{A(BC)}=-2\text{max} [0,\sum_j \lambda^{A(BC)}_{\text{neg},j}].
\end{equation}
with $\lambda^{A(BC)}_{\text{neg},j}$ the negative eigenvalues of the partially transposed density matrix describing the bipartition $A-(B,C)$. We can then determine the genuine tripartite entanglement content of the tripartite state by taking a geometric average of all the bipartitions, thus ensuring that the corresponding measure remains a true entanglement monotone. Therefore
\begin{equation}
\label{tripartneg}
{\cal N}_3= [{\cal N}_{A(BC)}{\cal N}_{B(AC)}{\cal N}_{C(AB)}]^{\frac{1}{3}}.
\end{equation}
Unfortunately, the extension of this measure to larger systems is not as straightforward and one has to resort to other tools for the quantitative analysis of multipartite entanglement. A rather handy one is embodied by $N$-concurrence~\cite{nconc}, which is the direct generalization of bipartite concurrence to the case of an {\it even} number of qubits. That is, one has to consider the eigenvalues $\eta_j$ of the $N$-spin flipped matrix
\begin{equation}
\rho(\otimes^N_{j=1}\hat\sigma^y_j)\rho^*(\otimes^N_{j=1}\hat\sigma^y_j)
\end{equation}
with $\rho$ the density matrix describing the state of the system and $\rho^*$ its complex conjugate. The $N$-concurrence $NC$ is then given by $NC=\max[0,\sqrt\eta_1-\sum^{N^2}_{j=2} \sqrt{\eta_j}]$ with $\eta_1\ge\eta_j~\forall{j}$. This measure is appealing due to its computationally handy form and its sensitivity to global entanglement. In fact, it is exactly zero if any qubit is separable from the rest of a system. Its non-nullity, though, is only a sufficient condition for multipartite entanglement as there are multipartite entangled states with vanishing $N$-concurrence, such as the N-qubit $W$-states.  According to the definition given by Wong and Christensen~\cite{WC}, in pure states N-concurrence quantifies the overlap between a given state and the one obtained upon application of an anti-unitary time-reversal operation. For qubit states, such operation is embodied by a collective spin-flip operation. $N$-concurrence is thus a natural choice when the state to study is the eigenstate of a Hamiltonian respecting time-reversal symmetry, such as those that can be written as the sum of tensor products of an even number of non-identity Pauli matrices~\cite{nconc}. The Ising Hamiltonian satisfies such condition when no magnetic field is applied, thus justifying our choice and our efforts here can well be seen as the quantification of the robustness of the ground state of a Ising-like model to the perturbation induced by the magnetic field.

There is no formal way to quantify the degree of non-locality of a given state. Nevertheless, the degree of violation of Bell-type inequalities can be taken as a an intuitive means of ascertaining the amount of non-locality~\cite{bell}. A more pragmatic viewpoint to quantify non-locality, on the other hand, is to consider the amount of noise required by a given state to stop violating a Bell-like inequality. Moreover, in the multipartite setting, a few ambiguities still remain in deciding which type of inequality to utilize. In fact, as very clearly pointed out by Cereceda~\cite{cereceda} by reprising a point originally made by Svetlichny~\cite{svetlichny}, $n$-particle entanglement is not physically equivalent to $n$-particle non-locality. In this respect, the violation of an $n$-particle Bell-like inequality of some sort by an $n$-particle entangled state is not enough, {\it per se}, to prove genuine multipartite non-locality.   When one moves into the multipartite setting Bell-type inequalities exist that probe various different types of correlation. Technically, this issue has produced considerable theoretical effort directed towards the formulation of appropriate Bell-like inequalities able to capture the genuine multipartite non-local nature of a given state. Noticeably, Svetlichny has derived an inequality for the tripartite case that, while obeyed by models assuming two-particle non-locality, is violated by quantum mechanical states that are genuinely three-particle entangled~\cite{svetlichny}. As discussed by Cereceda~\cite{cereceda}, a Svetlichny inequality (SI) is a righteous Bell inequality for the tripartite case and emerges as a valuable tool for the unambiguous assertion of the existence of genuine tripartite entanglement and tripartite non-locality in any three-particle state. To this task, the use of a standard Mermin's inequality~\cite{mermin} is not sufficient: on the contrary, quantum correlations that violate SI are strong enough to maximally violate Mermin's inequality as well. The inequality by Svetlichny has been independently extended to the $n$-partite scenario in Refs.~\cite{gisin,svet}. In particular, Collins {\it et al.}~\cite{gisin} have provided an iterative way to construct multipartite Bell inequalities for dichotomic observables. We shall call $o_j$ and $O_j$ the two outcomes of a local measurement performed over one of the particles. It should be noted, the explicit form of the observable can be dependent on the state to be tested. For instance, measuring along the incorrect axis for a particular state can lead to a failure to see any violation. In the proceeding section we will fix the appropriate observable for our model. Then, by setting $m_1=o_1$ ($M_1=O_1$), one constructs the polynomials
 \begin{equation}
 \label{theory}
 \begin{aligned}
m_n&=\frac{1}{2}m_{n-1}(o_n+O_n)+\frac{1}{2}M_{n-1}(o_n-O_n),\\
M_n&=\frac{1}{2}M_{n-1}(o_n+O_n)+\frac{1}{2}m_{n-1}(O_n-o_n).
\end{aligned}
 \end{equation}
Quantum mechanically, we shall interpret the polynomials $m_n$ and $M_n$ as sums of $n$-particle correlation functions for measurements having outcomes $o_j$ and $O_j$. We can then define the generalized Svetlichny polynomials 
\begin{equation}
\label{polys}
{\cal S}_n=\left\{
\begin{aligned}
&m_n~~~(n~\text{even})\\
&(m_n+M_n)/2~~~(n~\text{odd})
\end{aligned}
\right.
\end{equation}
The bound imposed to ${\cal S}_n$ by local hidden-variable models is $1$, while quantum mechanically an $n$-qubit GHZ state achieves the maximum value of $\sqrt{2^{n-1}}$ and $\sqrt{2^{n-2}}$ for an even and odd number of particles, respectively. Armed with such tools, we now tackle the main point of our study. 

\section{Behavior of non classical correlations: zero-temperature case}
\label{nonclassical}

\begin{figure}[t]
{\bf (a)}\hskip4cm{\bf (b)}
\psfig{figure=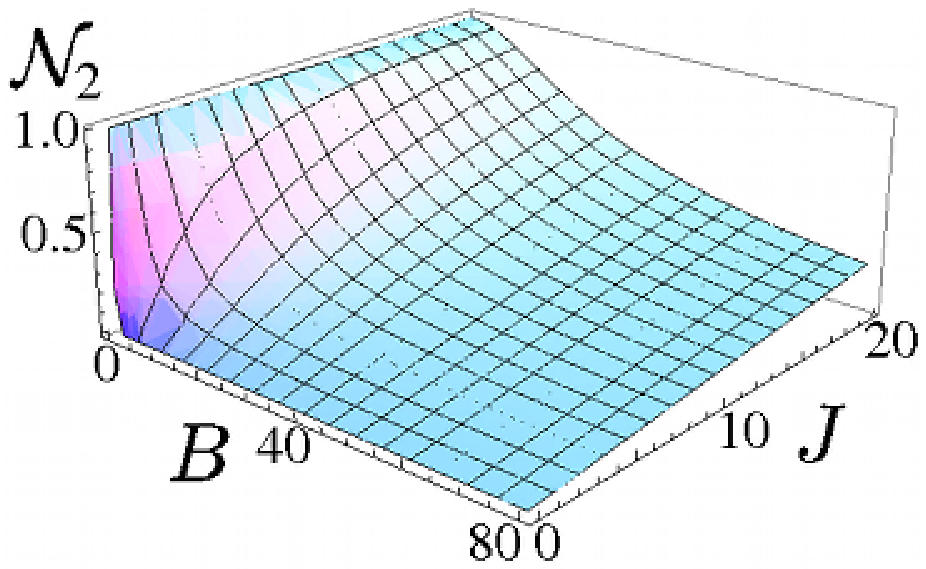,width=4cm,height=2.7cm}~~~\psfig{figure=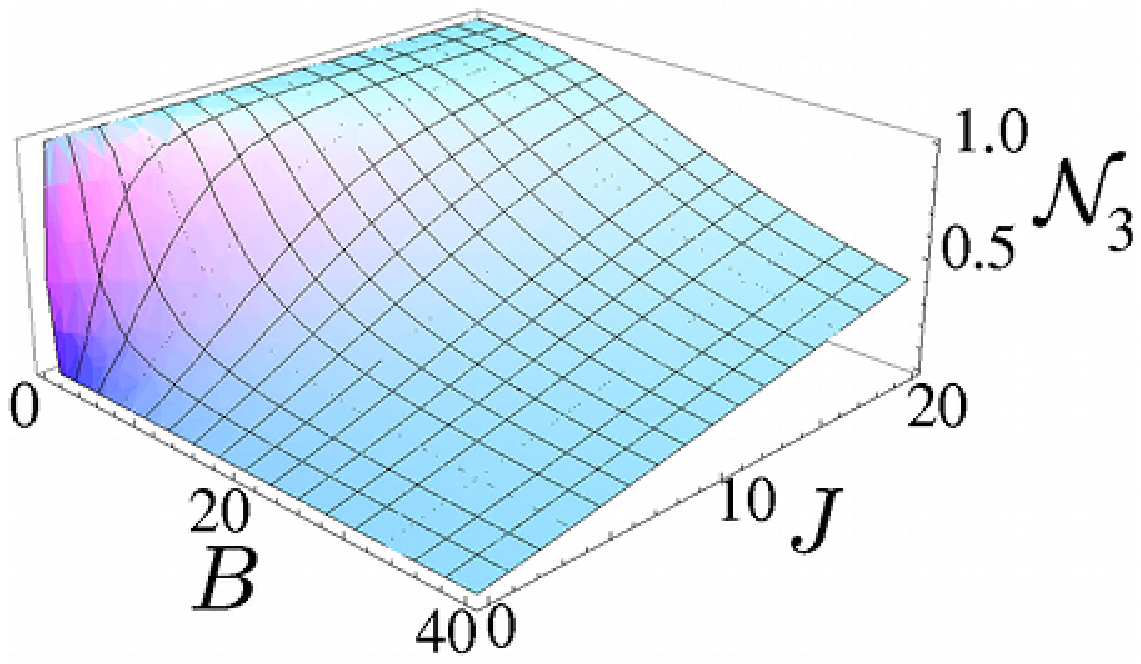,width=4.3cm,height=3.3cm}
\caption{(Color online). Behavior of entanglement against $B$ and $J$ for {\bf (a)} bipartite spin ring and {\bf (b)} tripartite spin ring. For small values of $B$ the ground state is highly entangled. As the field strength is increased we see the entanglement asymptotically approach zero.}
\label{zerotempentanglement}
\end{figure}

\begin{figure}[b]
\psfig{figure=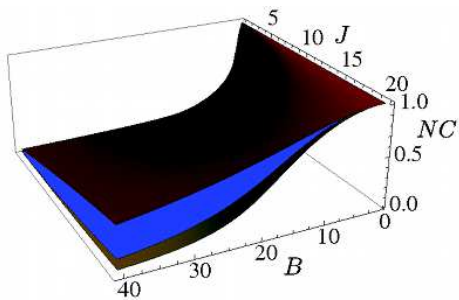,width=7cm,height=4cm}
\caption{(Color online). Comparison of the $N$-concurrence $NC$ corresponding to the ground state of a ring of $N=2,4$ and $6$ spins (curves from top to bottom one) at zero temperature. The $N$-concurrence is plotted against $B$ and $J$.}
\label{Nconc}
\end{figure}

We begin our analysis considering the dynamics of the correlations in the simplest case of just two coupled spins. By assuming that both the spins are at zero temperature, the ground state  of Eq.~(\ref{model}) has the form
\begin{equation}
\label{gs2}
\ket{\gs_2}=\frac{-(B+\Lambda_2/2)\ket{00}_{12}+J\ket{11}_{12}}{\sqrt{J^2+(B+\frac{\Lambda_2}{2})^2}}\end{equation}
with $\Lambda_2=-2\sqrt{B^2+J^2}\ge{B}$. Clearly, for $B\rightarrow{0}$, $\ket{\gs_2}\rightarrow(\ket{00}+\ket{11})/\sqrt 2\equiv\ket{\text{GHZ}^x_2}$, in agreement with the expectations for the ground state of the model at hand. The calculation of negativity is therefore straightforward and leads to 
\begin{equation}
{\cal N}_{2}=\frac{J}{\sqrt{B^2+J^2}},~~~~(B>0).
\end{equation}
The behavior against $J$ and $B$ is shown in Fig.~\ref{zerotempentanglement} {\bf (a)}. Clearly, as $B\rightarrow0$ the entanglement achieves a full ebit, while as we increase the amplitude of the magnetic field, the amount of zero-temperature entanglement decreases, vanishing asymptotically. This behavior is in agreement with what has been shown in Refs.~\cite{buzek,kendon}, where concurrence was considered.
We can extend our study to the case of tripartite entanglement. The zero-temperature tripartite negativity of the ground state of a three-spin ring can be calculated using Eq.~(\ref{tripartneg}) to be
\begin{equation}
{\cal N}_3=\frac{1}{3} \sqrt{\frac{3 J^2}{B^2-B J+J^2}-\frac{4 B-2 J}{\sqrt{B^2-B J+J^2}}+4}.
\end{equation}
Fig.~\ref{zerotempentanglement} {\bf (b)} shows its behavior. Analogously to the entanglement of the two-spin ring, we see that as the magnetic field is increased there is an asymptotic decay to zero. While at $B\rightarrow0$ the state is the maximally entangled GHZ state, interestingly as we increase the magnetic field strength we find the reduced state of any two spins becomes entangled, thus stating the departure of the ground state of the system from the GHZ class of entangled states. As discussed in Sec.~\ref{Sectools}, the case of an arbitrarily sized ring is not easily managed because of the lack of available measures for genuine multipartite entanglement. Yet, partial information on the distribution of quantum correlations within an even-parity length ring  can be effectively gathered by means of the $N$-concurrence. We have thus evaluated the $NC$ for the significant cases of $N=4$ and $6$ and compared it with the entanglement measured by Wootters concurrence evaluated for a two-spin chain. The results are shown in Fig.~\ref{Nconc} against $J$ and $B$ and reveal an increasing fragility of multipartite entanglement as $N$ grows. As it should be evident, at non-zero $J$ entanglement only asymptotically decreases as $B$ grows showing that, although the ground state of the ring progressively deviates from a GHZ state, a degree of multipartite entanglement is nevertheless kept by the spins.  

\begin{figure}[t]
{\bf (a)}\hskip3cm{\bf (b)}
\psfig{figure=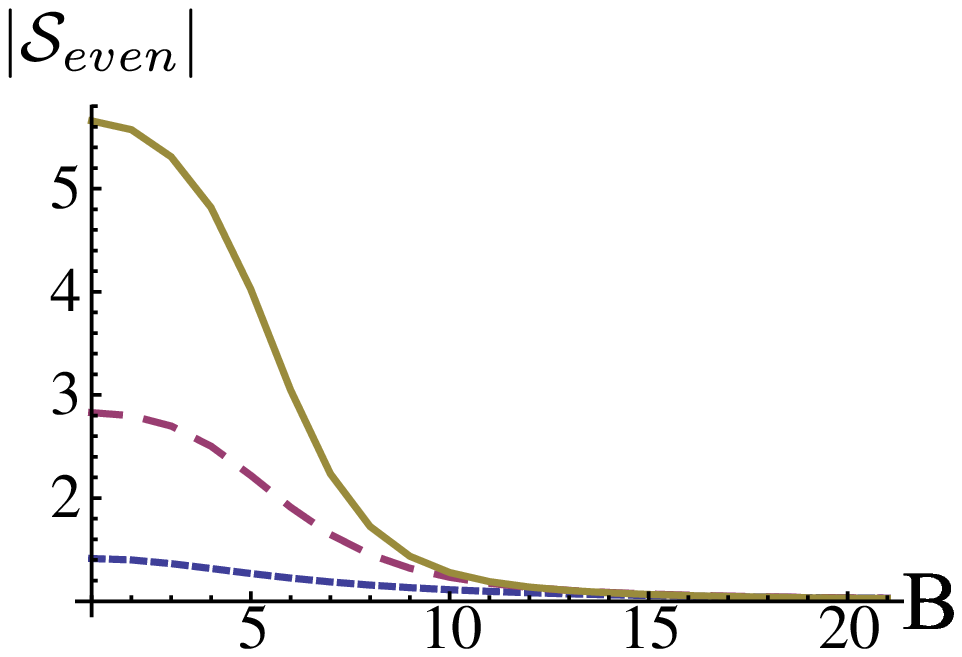,width=4.5cm,height=3.0cm}~\psfig{figure=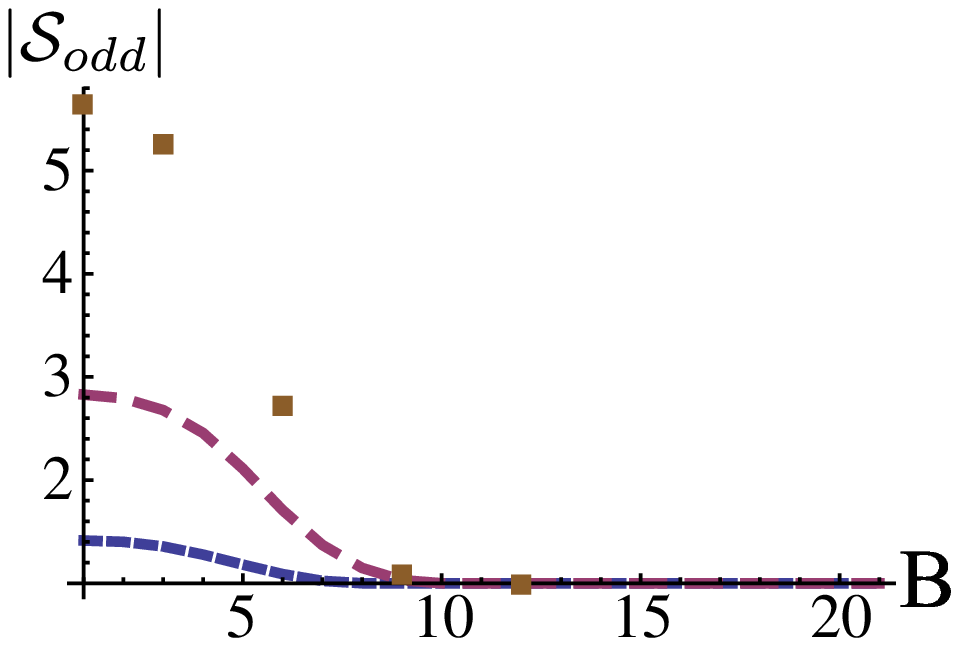,width=4.5cm,height=3cm}
\caption{(Color online). {\bf (a)} Multipartite non-locality of the ground state of a ring of an even number of spins against the amplitude of the magnetic field $B$ and at $J=5$. The solid line is for $N=6$, the dashed line for $N=4$ and, finally, the dotted one is for $N=2$. {\bf (b)} Same as in panel {\bf (a)} but for an odd number of spins. The dotted line is for $N=3$, the dashed for $N=5$ while the squares show the results for $N=7$.}
\label{zerotempnonlocality}
\end{figure}

However, the inability to discuss the non-classical features of correlations in the odd-parity scenario and the lack of a ``homogeneous" way to quantify entanglement (two different measures had to be used in the above) is certainly an unsatisfactory feature. Therefore, in order to provide a much more self-consistent characterization of the correlation properties within the ground states of the Ising ring, we move to the study of  non-locality. In such a two-spin setting, the well-known Closer-Horne-Shimony-Holt (CHSH) inequality~\cite{bell} can be used, which is encompassed in the formulation of the general multipartite problem given in Sec.~\ref{Sectools}. In order to run the CHSH inequality test, we thus project the state of the two-spin system onto the eigenstates of the local observable
\begin{equation}
R_i(\theta_i)=\left(
\begin{array}{cc}
 \cos\theta_i & -i\sin\theta_i \\
 i\sin\theta_i & -\cos\theta_i
\end{array}
\right)~~(i=1,2)
\end{equation}
and calculate the correlation function among the various combinations of outcomes. This is equivalent to 
\begin{equation}
C(\theta_1,\theta_2)=\bra{\gs_2}\hat{R}_1(\theta_1)\otimes\hat{R}_2(\theta_2) \ket{\gs_2}
\end{equation}
with $\theta_1$ and $\theta_2$ the angles defining the orientation of the observable $\hat{R}_i$. We then construct the polynomial $m_2$ by identifying each term appearing in the first of Eqs.~(\ref{theory}) with the correlation function obtained by using two different sets of angles, $(\theta_1,\theta_2)$ and $(\theta'_1,\theta'_2)$. We can then construct the CHSH function as
\begin{equation}
\label{CHSH}
{\cal S}_{2}=\frac{1}{2}[C(\theta_1,\theta_2)+C(\theta_1,\theta'_2)+C(\theta'_1,\theta_2)-C(\theta'_1,\theta'_2)].
\end{equation} 
Classically this function is bounded by $1$, while there are quantum mechanical states that violate this bound by up to a factor $\sqrt{2}$. In Fig~\ref{zerotempnonlocality} {\bf (a)} we see how a typical CHSH function $|{\cal S}_2|$ behaves against the amplitude of the transverse field $B$, for a set value of $J$. For nearly vanishing values of $B$ we have a maximally entangled state which maximally violates the bound set by local realistic theories. As the amplitude of $B$ increases, we see ${\cal S}_2$ asymptotically approaching the local realistic bound of $1$. 

Despite the apparent simplicity of the case studied above, the results achieved in the two-spin ring serve as useful milestones in the analysis to follow. In fact, we can extend our investigation to the genuinely multipartite scenario by using the information on the ground state of the Ising ring under scrutiny that can be gathered by means of the approach described in Sec.~\ref{Secmodel} and applying it to the formalism for multipartite Bell inequalities given in Sec.~\ref{Sectools}. Our approach goes as follows: although certainly possible, the calculation of the correlation function in terms of the fermionic operators putting  $\hat H$ into the form $\hat{H}_{\text{ff}}$ given in Eq.~(\ref{free}) is not straightforward, especially under the scalability viewpoint. Indeed, as significantly remarked by Osborne and Nielsen~\cite{osborne}, calculating correlation functions for more than two particles is, in general, a daunting problem. One should, in fact, work out the expression of $\hat{R}_{i}(\theta_i)$ in the fermionic representation, keep in consideration the anti-commutation rules obeyed by fermions and re-cast the tensor product of $N$-particle local observables in terms of the operators $\hat{b}_k$ and $\hat{b}^\dag_{k}$. Expectation values over the ground state $\ket{\gs_N}$ of an $N$-spin ring could be finally evaluated. Instead, we decided to use the information on the energy $\Lambda_N$ obtained by means of the apparatus set up in Sec.~\ref{Secmodel} so as to determine the form of the ground state of the system as expressed in terms of the elements of the computational basis. This is easily and computationally very conveniently done by solving the set of equations 
\begin{equation}
\label{eig}
(\hat{H}-\Lambda_N\openone_{2^N})\ket{gs_N}={\bm 0}, 
\end{equation}
where $\Lambda_N$ is now a known quantity, $\hat\openone$ is the identity matrix and ${\bm 0}$ is the identically null vector. Once the structure of the ground state is determined, the evaluation of the necessary correlation functions is largely simplified. To the best of our knowledge, our study is the first to address multi-spin correlations over the ground state if an Ising ring. We have thus studied the behavior of the Svetlichny function against the amplitude of the magnetic field $B$ at a set value of $J$ (we remind that the important parameter, in this problem, is the ratio $J/B$). In Figs.~\ref{zerotempnonlocality} {\bf (a)} and {\bf (b)} we show the results of our investigation for rings of up to $7$ spins. It is worth stressing that larger rings can well be studied by our method. Given the determination of the ground state has been reduced to the solution of the eigen-equation Eq.~(\ref{eig}) we require only the solution of $2^N$ linear equations. It is in the evaluation of the multipoint correlation functions that care must be taken. Here the number of correlation functions to be evaluated scales with $2^N$, and we must then optimize the generalized Svetlichny polynomials Eq. (\ref{polys}) with respect to $2N$ parameters. This leads to escalating computational cost making rings larger than $N=7$ intractable. We have split the behavior of the ${\cal S}_{n}$ functions against the parity of the length of a ring, thus defining ${\cal S}_{even}$ and ${\cal S}_{odd}$.  As it will be shown later, parity appears to be relevant in the investigation of the features here under scrutiny. 

The dependence of the the two parity-related Svetlichny functions on $B$ is mutually fairly similar and overall consistent with the behavior revealed for ${\cal S}_{2}$: at large values of $J/B$, the ground state of the ring exhibits strong non-locality, up to the maximum allowed value set by the number of spins being considered. As $B$ grows (for a set value of $J$), the non-local nature of $\ket{\gs_n}$ deteriorates. Remarkably, though, while the even-parity case approaches the local realistic bound only asymptotically with $B$, a finite value of its amplitude is sufficient to set $|{\cal S}_{odd}|=1$ exactly. Although it is difficult to establish an explicit relation at this level, we believe that such a parity-induced effect is related to the {\it differences} in the way the ground state of an Ising ring has to be considered in the even- and odd-number of spins cases, as highlighted in Sec.~\ref{Sectools} when the ring Hamiltonian was diagonalized. However, it should be remarked that longer chains of an odd number of spins  require a larger $B$ to reach the local realistic bound, while in a long chain, the addition of a single spin would make a very small difference, as it is intuitively reasonable to expect. Therefore, one can extrapolate to the thermodynamical limit the behavior of non-locality and claim that the parity effect highlighted above would eventually disappear. As a side remark, the reduced state of any two-spin subsystem, regardless of the length of a ring, despite being entangled do not violate Eq.~(\ref{CHSH}).  

\begin{figure}[b]
{\bf (a)}\hskip3.0cm{\bf (b)}\hskip3.0cm{\bf (c)}
\psfig{figure=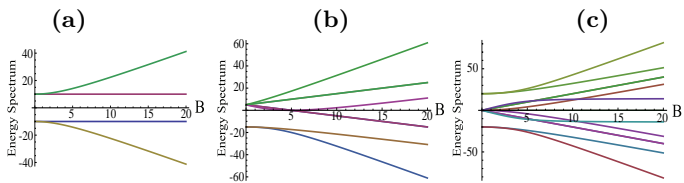,width=9.0cm,height=2.0cm}
\caption{(Color online). Energy spectrums for a spin ring of $N=2$ [panel {\bf (a)}], $3$ [panel {\bf (b)}] and $4$ [panel {\bf (c)}] against $B$. We have taken $J=5$.}
\label{spectrums}
\end{figure}

\begin{figure*}[t]
{\bf (a)}\hskip3cm{\bf (b)}\hskip4cm{\bf (c)}
\psfig{figure=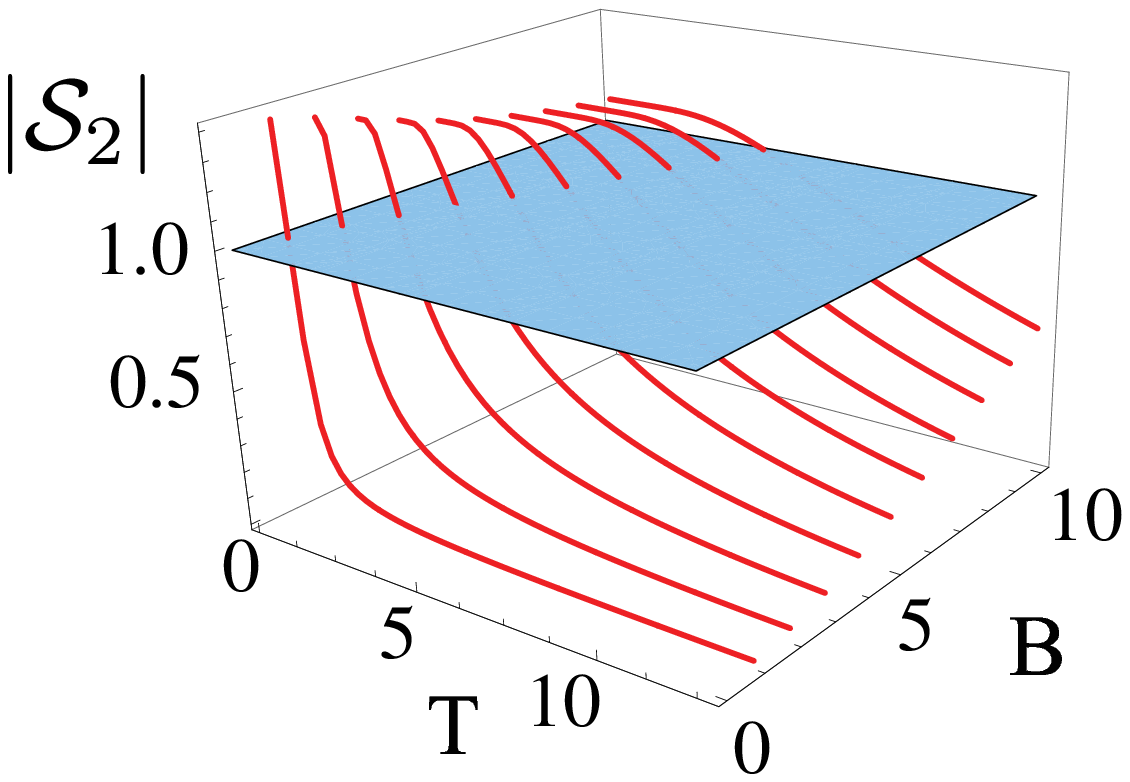,width=4.7cm,height=3.5cm}~~\psfig{figure=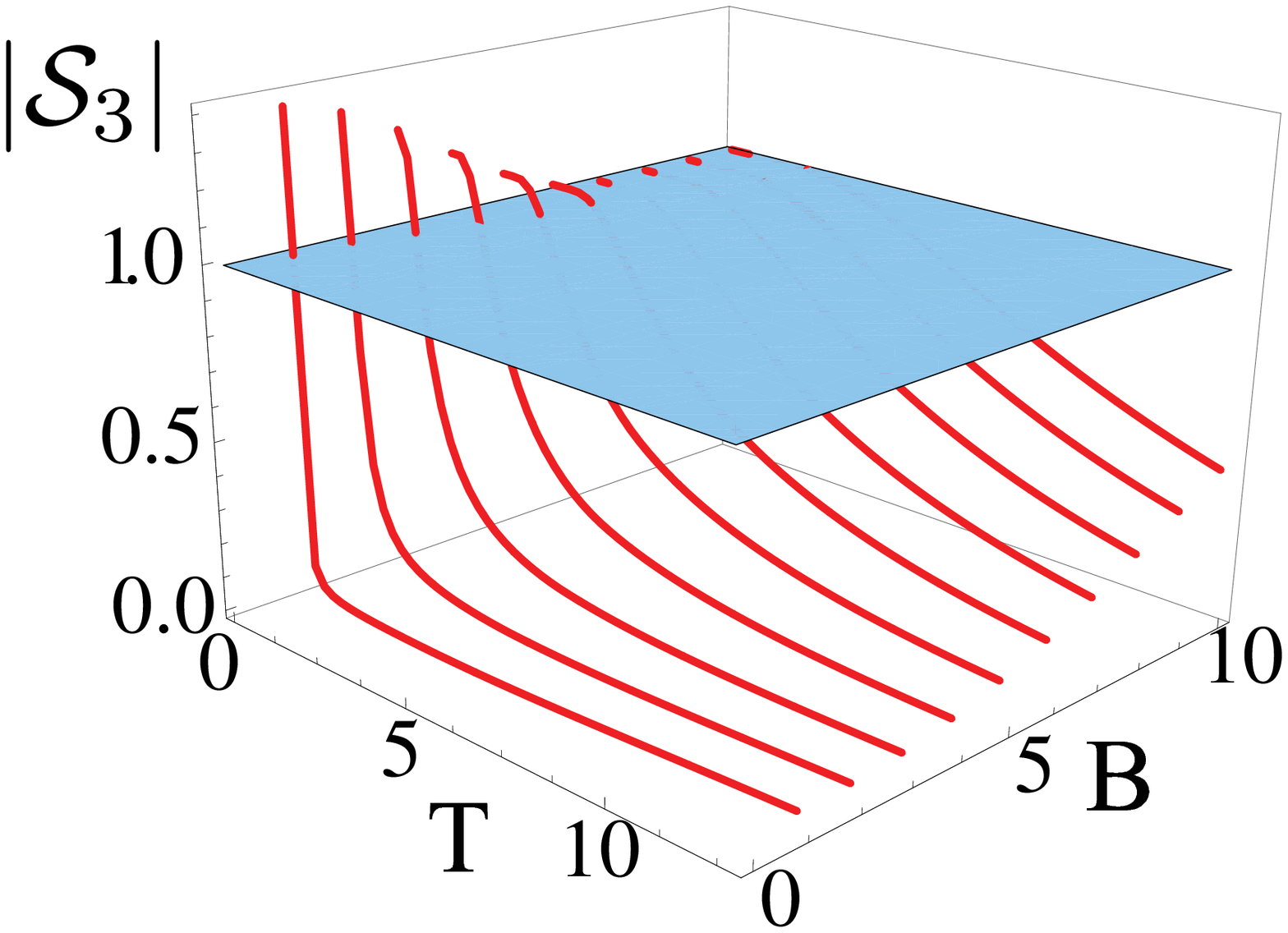,width=4.5cm,height=3.6cm}~~\psfig{figure=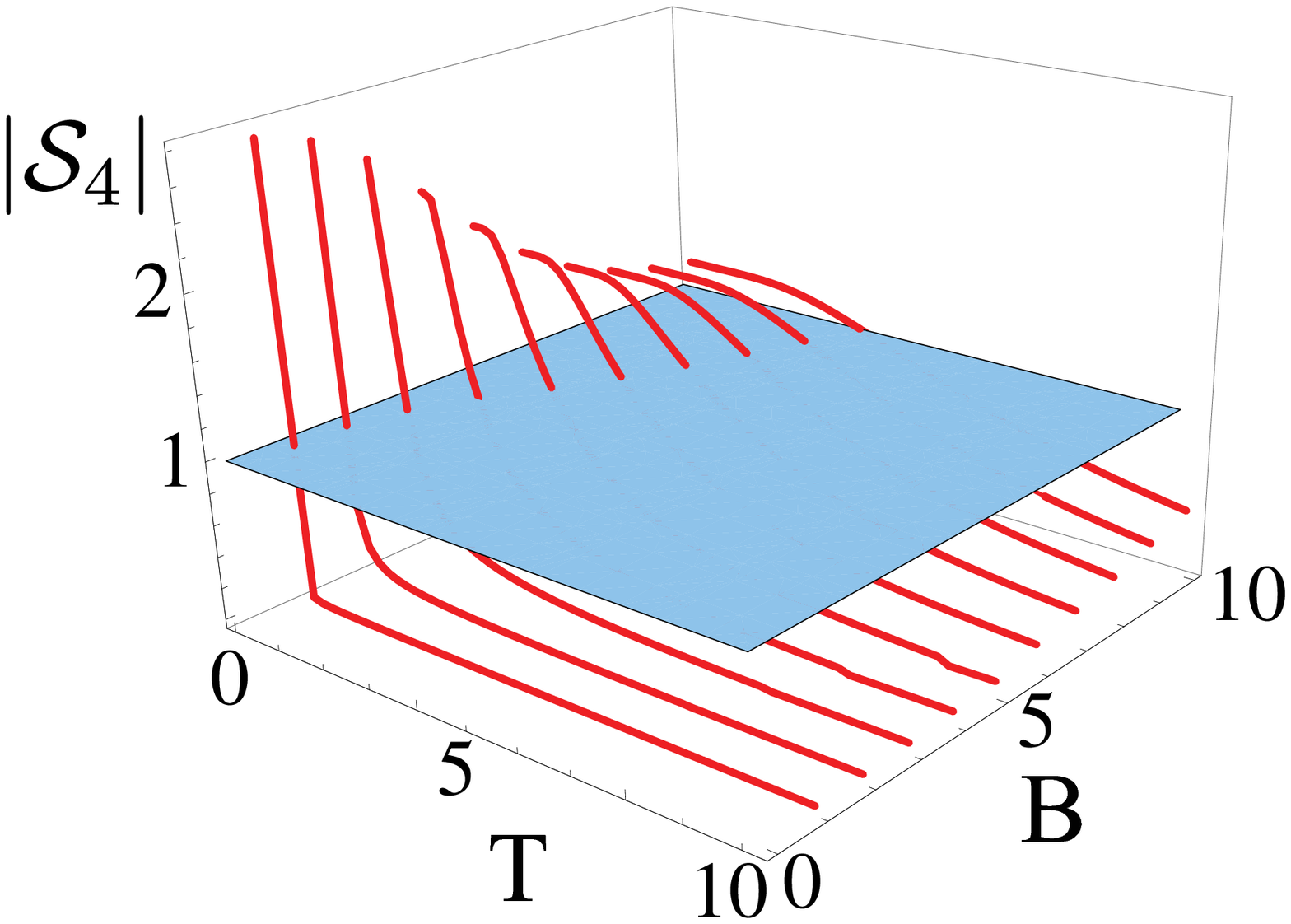,width=4.6cm,height=3.5cm}
\caption{(Color online). Behavior of Svetlichny functions ${\cal S}_{N}$ for the state of a ring with $J=5$ and $T\neq0$. We address the case of $N=2,3$ and $4$ (panel ${\bf (a)}$, ${\bf (b)}$ and ${\bf (c)}$ respectively) as representative of the general trend exhibited by arbitrary $N$ values. Each curve in the same plot is for for a specific choice of $B\in[1,10]$. Regardless of $N$, as $B$ increases the degree of violation of the appropriate Bell-like inequality decreases. However, the resilience to spoiling effects due to temperature increases.}
\label{thermaleffects}
\end{figure*}

\section{Behavior of non classical correlations: thermal effects}
\label{TnonZero}

We now investigate the behavior of non-classical correlations in the non-zero temperature case. Using Eq.~(\ref{thermalstate}) we can determine the thermal state of the Hamiltonian. In general, this necessitates of the full spectrum of eigenstates the Hamiltonian, whose determination for a large number of spins becomes a very difficult task to achieve. However, very useful information comes from the study of the energy spectrum of $\hat H$. Fig.~\ref{spectrums} shows the spectrum of a ring of  two, three and four spins for $J=5$ and against increasing field $B$~\cite{nota}. It is seen that the splitting between the two lowest energy states increases with the strength of the magnetic field. In the region of  $B\ll J$, the ground and first excited energy levels are almost degenerate and a much larger energy gap separates the quasi-degenerate doublet from the rest of the spectrum. This implies that an excellent approximation of the thermal state of the ring comes from considering an admixture of $\ket{\gs_N}$ and $\ket{\text{first}_N}{=}\hat{b}^\dag_k\ket{\gs_N}$ with coefficients determined by the Boltzmann factor corresponding to the energies $\Lambda_N$ and $\Lambda^1_N{=}\bra{\text{first}_N}\hat{H}_{\text{ff}}\ket{\text{first}_N}$ such as
\begin{equation}
\label{appro}
\varrho(T){\simeq}\left(e^{-\frac{\beta\Lambda_N}{T}}\ket{\gs_N}\bra{\gs_N}+e^{-\frac{\beta\Lambda^1_N}{T}}\ket{\text{first}_N}\bra{\text{first}_N}\right)/{\cal Z}.
\end{equation}
Clearly, the approximation is legitimate as long as the state~(\ref{appro}) remains quasi-normalized, that is $(\exp[{-\frac{\beta\Lambda_N}{T}}]+\exp[{-\frac{\beta\Lambda^1_N}{T}}])/{\cal Z}\simeq{1}$. The proximity of the first two energy levels for small values of $B$ implies that even a small amount of thermal energy is enough to fully mix the two lowest eigenstates, which might well result in the wash-out of the quantum correlations within the system. On the other hand, for increasing values of $B$, a larger temperature is required to mix the eigenstates. 

We have examined the entanglement properties for even values of $N$ {\it and} $N=3$ (which, we remark, is the only odd case treatable through tripartite negativity) against the temperature. For $B\ll J$, entanglement suddenly disappears at finite values of $T$. On the other hand, for $B\approx J$, quantum correlations appear to cope much better against the effects of a non-zero temperature: as $B$ gets larger, the state of the ring is able to withstand larger values of $T$ before we starting to decay. Yet, this effect is accompanied by a reduction in the amount of entanglement shared by spins.  Again, in order to bypass the problems inherent in entanglement measures for mixed multipartite states, we use the tools provided by multipartite non-locality tests. Figs.~\ref{thermaleffects} {\bf (a)-(c)} show the behavior of the generalized Svetlichny functions for a few significant cases. Evidently, as $B$ grows, while paying in terms of degree of violation of the appropriate Bell-like inequality, we significantly enlarge the range of $T$ within which the correlations within the state of a ring cannot be described by a local realistic theory. 

\begin{figure}[b]
\psfig{figure=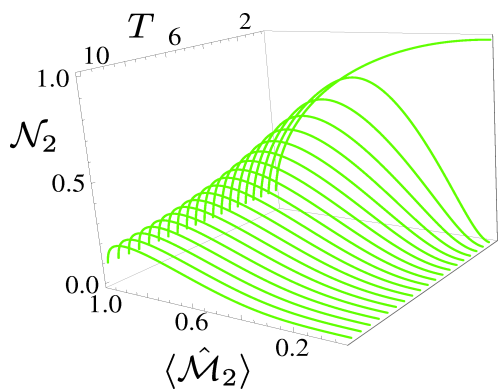,width=6.5cm,height=5cm}
\caption{(Color online). Two-spin negativity against magnetization and temperature for $J=5$. We have taken $T=1,..,20$ (increasing at unit steps). The entanglement for magnetization going toward $1$ stays at finite non-zero values only because we have considered $B\in[0,100]$. By enlarging this range, smaller final values of negativity would be reached. In fact, we have ${\cal N}_2\rightarrow0$ as $\langle\hat{M}_2\rangle\rightarrow1$.}
\label{fig5}
\end{figure}

This {\it protection effect} can be understood as follows: the increasing magnetic field tends to align the spins in the ring, thus affecting the strength of the spin-spin correlations. On the other hand, the tendency to mutual alignment of the spins makes them less sensitive to the fluctuations induced by thermal effects. At the same time, a growing $B$ makes the spectral gap between ground and first excited state of the ring wider, thus making the admixture between the ground state and the rest of the spectrum less likely to occur. {We can link the effects analyzed above to the magnetization of the ring so as to provide a much neater connection between the correlation properties of the system and its intrinsic macroscopic features, which is certainly a very interesting point to assess. Let us concentrate on the two-spin ring case, which is non-trivial enough to encompass the general features we are interested in.  We consider the expectation value of the global magnetization operator $\hat{\cal M}_2=\sum^2_{j=1}\hat{\sigma}^z_j$ over the state of a two-spin ring at temperature $T$. Clearly, the modulus of this function grows monotonically from zero to one at a rate slowed down by the increasing temperature. Here, we are interested in the connection between non-classical correlations and the global magnetization $\langle\hat{\cal M}_2\rangle$. Due to the dimension of the problem at hand, we can equally use negativity or the Bell-like function ${\cal S}_2$ and we focus on the former simply for easiness of calculations. In Fig.~\ref{fig5} we plot ${\cal N}_2$ against $\langle\hat{\cal M}_2\rangle$ for a range of values of the temperature. At $T=0$, small values of the magnetization (corresponding to large values of the ratio $J/B$) correspond to a very large degree of entanglement, in agreement with the analysis so far~\cite{buzek}. A growing magnetization sees ${\cal N}_2$ decreasing to zero. The behavior changes abruptly for even small non-zero values of $T$, when  we start from an initially maximally entangled state and as we increase the magnetization of the system we see a decrease in non-classical correlations. For the bipartite case we see the correlations become purely classical only when the system is fully saturated by the magnetization, i.e. when $M=1$ or equivalently $B\rightarrow\infty$. Qualitatively the results hold in the higher dimensional cases. While the study of this figure against temperature reinforces the features revealed above (entanglement decays more slowly at larger values of $B$, although the maximum achievable degree of negativity decreases), we can also infer that larger values of the magnetic field are required in order to optimize the entanglement within the ring, if $T$ grows. This is a clear manifestation of the {\it correlation-by-alignment} effect discussed before: the thermal randomness is counteracted by the order induced by a strong magnetic field. However, if the alignment induced by $B$ exceeds the threshold determined by the degree of thermal disorder, quantum correlations start to vanish as the spins tend to be only classically correlated (as expected in an Ising ring beyond the critical point). A qualitatively similar analysis holds for longer rings. By controlling the macroscopic features of magnetization, one could thus determine the degree of quantum correlations set among the elements of the ring. 

\section{conclusions}
\label{conclus}
We have studied the non-classical correlations naturally occurring in the ground and thermal state of an Ising ring in transverse magnetic field. The focus of our analysis was the behavior of multipartite entanglement and non-locality as quantified by the violation of {\it ad hoc} many-spin Bell-like inequalities~\cite{gisin}. At zero temperature, while even-length rings cannot be described by local realistic theories even at the limit of asymptotically infinite strength of the magnetic field, odd-length ones behave differently: there is always a finite choice of $B$ beyond which the bound imposed by local realism is satisfied. We have been able to exactly tackle multipartite correlation functions in such a many-body problem, therefore going significantly far from the standard analysis of the behavior of quantum correlations in interacting many-body models. At non-zero temperature, we have found that the thermal randomness that would destroy the non-classical correlations found in the ground state of the model can be counteracted simply by setting a large enough magnetization of the ring: a large magnetic field, in fact, while decreasing the degree of violation of Bell-like inequalities, will make the generalized Svetlichny functions used for non-locality test resilient against thermal spoiling effects. This suggests that  a genuinely multipartite quantum-correlated state in an Ising ring can be generated simply by controlling the macroscopic properties of the system, i.e. magnetization and temperature. We believe our results will trigger further interest in examining entanglement and non-locality in many-body interacting systems.

\acknowledgments
We thank Dr. Nicolas Brunner, Dr. Tony Apollaro, Prof. Sougato Bose and Prof. Vlatko Vedral for invaluable discussions. We acknowledge financial support from DEL and the UK EPSRC (EP/G004579/1),

\end{document}